\pdfoutput=1
\PassOptionsToPackage{table}{xcolor}
\documentclass[11pt,xcolor=table]{article}

\usepackage[]{acl}

\usepackage[table]{xcolor}
\usepackage{times}
\usepackage{latexsym}
\usepackage{graphicx}
\usepackage{tabularx}
\usepackage{booktabs}
\usepackage{multirow}
\graphicspath{ {./images/} }

\defcitealias{principles-to-actions}{NCTM, 2014}

\usepackage[T1]{fontenc}
\usepackage[utf8]{inputenc}

\usepackage{microtype}

\newcommand\unfiltered{\textsc{unfiltered}}
\newcommand\filtered{\textsc{filtered}}

\title{Computationally Identifying Funneling and Focusing Questions in Classroom Discourse}

\author{\vspace{.5em}{\bf Sterling Alic}\textsuperscript{1}\quad {\bf Dorottya Demszky}\textsuperscript{1}\quad {\bf Zid Mancenido}\textsuperscript{2}\quad {\bf Jing Liu}\textsuperscript{3}\\ \vspace{.5em} {\bf Heather Hill}\textsuperscript{2}\quad {\bf Dan Jurafsky}\textsuperscript{1}\\ \vspace{.5em} \textsuperscript{1}Stanford University\quad \textsuperscript{2}Harvard University\quad  \textsuperscript{3}University of Maryland\\ \texttt{\{salic, ddemszky\}@stanford.edu}}

\begin{document}
\maketitle
\begin{abstract}
Responsive teaching is a highly effective strategy that promotes student learning. In math classrooms, teachers might \emph{funnel} students towards a normative answer or \emph{focus} students to reflect on their own thinking, deepening their understanding of math concepts. When teachers focus, they treat students' contributions as resources for collective sensemaking, and thereby significantly improve students' achievement and confidence in mathematics. We propose the task of computationally detecting funneling and focusing questions in classroom discourse. We do so by creating and releasing an annotated dataset of 2,348 teacher utterances labeled for funneling and focusing questions, or neither. We introduce supervised and unsupervised approaches to differentiating these questions. Our best model, a supervised RoBERTa model fine-tuned on our dataset, has a strong linear correlation of $.76$ with human expert labels and with positive educational outcomes, including math instruction quality and student achievement, showing the model’s potential for use in automated teacher feedback tools. Our unsupervised measures show significant but weaker correlations with human labels and outcomes, and they highlight interesting linguistic patterns of funneling and focusing questions. The high performance of the supervised measure indicates its promise for supporting teachers in their instruction.\footnote{Data and code are available at \url{https://github.com/sterlingalic/funneling-focusing}}
\end{abstract}

\section{Introduction}

Students are more engaged and learn more when teachers pose carefully chosen questions to draw out student thinking, and then attend closely to what students say \cite{BLAZAR201516, questioning-patterns}. One way that teachers do this is by using focusing question patterns; i.e., “attending to what the students are thinking, pressing them to communicate their thoughts clearly, and expecting them to reflect on their thoughts and those of their classmates”
\citep[National Council][hereafter NCTM]{principles-to-actions}. Focusing is often contrasted to the less effective yet more common question pattern of funneling, where teachers pose “a set of questions to lead students to a desired procedure or conclusion, while giving limited attention to student responses that veer from the desired path” \citepalias{principles-to-actions}. The use of focusing questioning patterns has been linked to better student learning outcomes and confidence in mathematics \citep{hagenah2018funneling,franke2001learning}.

\begin{figure}[t!]
    \includegraphics[width=\linewidth]{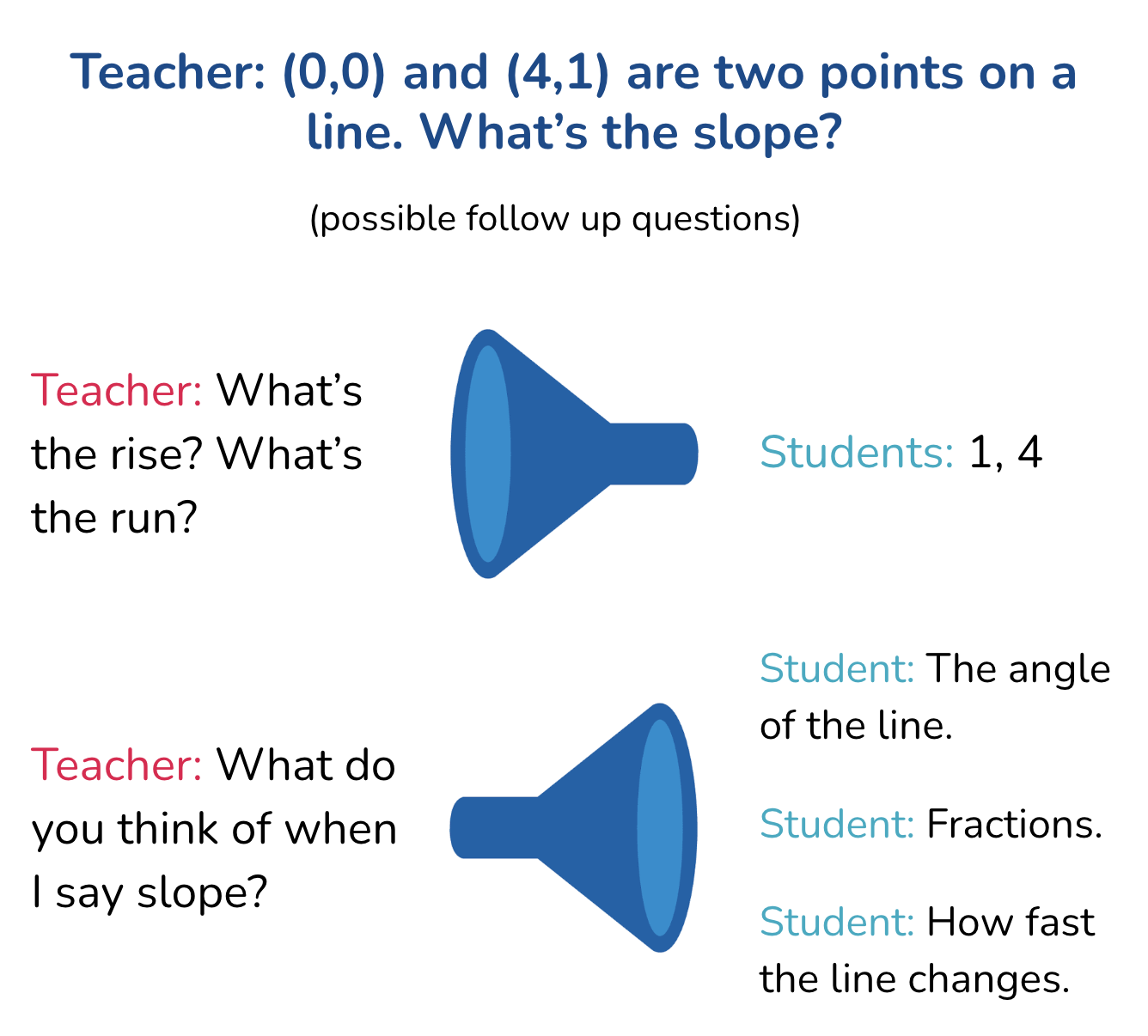}
    \caption{Example teacher utterance and possible student replies, illustrating the difference in funneling (top exchange) and focusing (bottom exchange). \cite{questioning-patterns} \label{fig:funnel_focus_example}}
\end{figure}

Supporting teachers to use more focusing question patterns requires first helping them to identify the extent to which they are focusing or funneling in their own classrooms. However, the current methods of measuring funneling and focusing are resource intensive, requiring manual classroom observation \citep[e.g.,][]{hagenah2018funneling}. Developing computational methods for identifying funneling and focusing thus present an opportunity to provide automated feedback for questioning patterns at scale. Recent tools that provide automated feedback to teachers on discourse moves have been effective at improving their uptake of student contributions and student outcomes \citep{demszky2021can} and helped raise awareness about different instructional talk moves \citep{jacobs2022promoting}. One promising application of an automated measure of questioning patterns is to build a similar tool that encourages teachers to engage with their students by asking more focusing questions.

We propose several approaches for computationally identifying funneling and focusing questions, including supervised and unsupervised modeling. In order to develop our approaches, we create a dataset of 2,348 student-teacher exchanges sampled from elementary math classroom transcripts, each annotated by three domain experts for teachers’ use of funneling and focusing questions, or neither. Then, we fine-tune a supervised RoBERTa  model \citep{liu2019roberta} on the annotated data. This model has the highest correlation of $.76$ with human judgments, among our proposed models.

We also explore several unsupervised learning approaches, in order to encourage domain-transferability, to account for the lack of labeled data in most educational settings, and to analyze the linguistic patterns that drive funneling vs focusing questions. Our first unsupervised model hinges on the assumption that the range of possible student responses are narrower for funneling questions than for focusing ones. In Figure~\ref{fig:funnel_focus_example}, we see that the teacher's funneling questions about the rise and run are quantitative in nature, so we can more confidently predict that the student response will be a number. Conversely, focusing utterances, which encourages students to reflect on their own thinking, tend to have a wider range of valid responses. The teacher's focusing question in Figure~\ref{fig:funnel_focus_example} shows that the students can think about the slope in many different ways, so we can less confidently predict what the student reply will be. Following this intuition, we adapt \citet{zhang-danescu-niculescu-mizil-2020-balancing}'s measure of \emph{forwards-range}, an unsupervised measure that quantifies the strength of our expectation of a reply to a given utterance.

We also use other linguistic features informed by educational theory as measures to identify funneling and focusing. Since focusing examples probe student thinking and understanding, we use the count of \emph{cognitive verbs} present in an utterance as one of the features. \emph{Question words and phrases} also provide insight into classifying closed-ended and open-ended questions, so we include both the count of unigram and bigram question words as features. Table~\ref{tab:feature_table} shows the list of words we used for each feature. We find that while some of these features correlate significantly with human judgments (e.g. forwards-range and the use of ``why''), these correlations are significantly weaker than those of the RoBERTa model.

To further validate our measures and to understand the link between funneling and focusing and educational outcomes, we correlate our measures with observation scores of instruction quality and student engagement and with value-added scores. Value-added scores are statistical estimates of a teacher’s contribution to student test score growth, which serve important indicator of student learning and achievement. We find that our RoBERTa model correlates strongly with all of these outcomes, which is a significant finding in the context of educational measurement \citep{kraft2020interpreting}, and it indicates the promise of this measure to support teachers and students.

\section{Contributions}
We make the following contributions in this paper.

\begin{enumerate}
    \item We propose the task of identifying funneling and focusing questions in classroom discourse.
    \item We create and release an annotated dataset of 2,348 teacher turns labeled for funneling or focusing questions or neither.
    \item We propose supervised and unsupervised approaches to identify funneling and focusing questions. Our unsupervised approaches include counting lexical features (e.g. question words and cognitive verbs) and estimating the expected diversity of responses to a teacher utterance. Our best-performing approach, a RoBERTa model, has a correlation of $.761$ with human annotations.
    \item We show that our estimates of funneling and focusing have a significant positive correlation with meaningful educational outcomes related to instruction quality and student achievement.
\end{enumerate}

\section{Related Work}
Many researchers have measured the types of question patterns that teachers use in classrooms by hiring and training raters to manually code transcripts of teacher-student discourse \cite{boaler2004importance,kane2015national,gregory2017my}. While this measurement approach has been useful for identifying effective teaching practice in well-funded large-scale research studies, it is too costly to be scalable.

Computational methods for measuring question patterns in classrooms offer both the potential to undertake more research in this area, as well as the potential to support teachers to improve their classroom practice by automatically coding aspects of their classroom discourse for them to review.

Prior work in computationally analyzing classroom discourse has employed a variety of techniques to automatically detect teacher discourse variables. Recent advances in natural language processing has led to a larger presence of work applying neural methods with varying levels of success in detecting classroom discourse variables, such as semantic content, instructional talk, and elaborated evaluation \cite{jensen-transfer-learning-discourse, song-semantic-dialogue}. For unsupervised approaches, \citet{demszky2021measuring}, which is also most similar to our work in terms of approach and dataset, propose an unsupervised measure of teachers' uptake of students’ contributions, and we use their sample in our annotation for funneling and focusing. Other computational work on questions in classroom discourse has focused on detecting questions in live classroom audio \citep{automatic-question-detection,blanchard2016identifying} and measuring the authenticity of questions in classroom discourse \citep{cook2018open,kelly2018automatically}. Our task closely relates to the task of detecting authentic questions but instead of using the CLASS framework used by prior work, we draw on the math education literature to develop our own coding instrument for funneling and focusing. In addition, while prior work in computationally analyzing questions uses feature-based classification, we also apply state-of-the-art neural machine learning models to solve this task. 

Our proposed task of identifying funneling and focusing questions is situated among related dialogue tasks where the goal is to predict a label for a set of turns in dialogue. General approaches to this task have employed supervised classifiers in a variety of settings, such as to classify sarcasm in social media dialogue and participant roles in cyberbullying \cite{lukin-sarcasm-nasty, jacobs_van_hee_cyberbullying_2022}. Similar to our approach of identifying patterns that generalize beyond annotated data, others in this domain have also  found meaningful patterns and features in labeled data that successfully generalized to unlabeled data \cite{oraby-etal-2015-thats, oraby-etal-2015-emotional-factual}. 

Our work is also closely related to the computational study of conversations. We build on \citet{zhang-danescu-niculescu-mizil-2020-balancing}'s unsupervised measure of forwards-range, which was originally developed to analyze strategies in counseling conversations. \\
\section{Dataset}
We create a new open-source dataset labeled for funneling and focusing questions with the help of domain experts. We recruit former and current math teachers and educators trained in classroom observation to annotate 2,348 examples of teacher-student exchanges. We use the same sample of exchanges as \citet{demszky2021measuring} --- they are sampled from transcripts of 45-60 minute long 4th and 5th grade elementary math classroom observations collected by the National Center for Teacher Effectiveness (NCTE) between 2010-2013 \citep{kane2015national}.\footnote{The only difference between our sample and that of \citet{demszky2021measuring} is that we include an additional 102 examples that were rated by all 13 raters, instead of only the examples rated by 3 raters. } The transcripts represent data from 317 teachers across 4 school districts in New England that serve largely low-income, historically marginalized students. Transcripts are fully anonymized: student and teacher names are replaced with terms like ``Student'', ``Teacher'' or ``Mrs. H''.\footnote{Parents and teachers gave consent for the study (Harvard IRB \#17768), and for de-identified data to be retained and used in future research. The transcripts were anonymized at the time they were created.}

\subsection{Annotation}

Our annotation framework for funneling vs focusing is designed by experts in math quality instruction, including our collaborators, math teachers and raters for the Mathematical Quality Instruction (MQI) coding instrument, used to assess math instruction \citep{learning2011measuring}. We prepare a dataset of utterance pairs $(S,T)$ for annotation, where $S$ is a student utterance and $T$ is a subsequent teacher utterance following the approach of \citet{demszky2021measuring}.  In the annotation interface, raters can see the utterance pair $(S,T)$, the lesson topic, which is manually labeled as part of the original dataset, and two utterances immediately preceding $(S,T)$ for context.

A teacher utterance needs to meet three criteria in order to be categorized as funneling vs focusing: it needs to (i) relate to math, (ii) follow up on the previous student utterance, (iii) include a question. For example, a question such as ``Can you sit down please?'' cannot be classified as funneling or focusing because it does not relate to math.  Similarly, if the teacher asks a question on a new topic, their question cannot be rated for funneling vs focusing, since it does not follow up on the previous student utterance. Therefore, annotators are first asked if a teacher utterance meets these three criteria. If so, raters are asked to indicate whether the utterance can be categorized as funneling or focusing. The coding protocol is included among the supplementary materials.

We recruited expert raters (with experience in teaching and classroom observation) whose demographics were representative of US K-12 teacher population. We followed standard practices in education for rater training and calibration. We conducted several pilot annotation rounds (5+ rounds with a subset of raters, 2 rounds involving all 13 raters), quizzes for raters, thorough documentation with examples, and meetings with all raters. After training raters, we randomly assign each example to three raters. Table~\ref{tab:annotated_examples} includes a sample of our annotated data, with majority rater judgments.

\paragraph{Post-processing.} We create two datasets to separately measure our methods' ability to identify funneling and focusing questions in naturally occurring data --- i.e. including data that does not meet the criteria above --- and its ability to separate funneling questions from focusing ones. We create a dataset called \unfiltered{}, where we replace raters' judgments for a teacher utterance not meeting the criteria above with $0$, funneling with $1$ and focusing with $2$. We also create a dataset called \filtered{}, where we replace raters' judgments for a teacher utterance not meeting the criteria above with \texttt{NaN}, funneling with $0$ and focusing with $1$. Then, we z-score each raters' judgments, and compute the average of z-scores across raters to obtain a single label for each example in each dataset. This process yields 2348 unique examples for \unfiltered{} and 1566 unique examples for \filtered{}.

\paragraph{Rater agreement.}  We obtain an average interrater leave-out Spearman correlation of $\rho=.644$ for \unfiltered{} (Fleiss $\kappa=.415$\footnote{We prefer to use correlations because kappa has undesirable properties \citep[see][]{delgado2019cohen} and correlations are more interpretable and directly comparable to our models' results (see later sections).}), and $\rho=.318$ for \filtered{} (Fleiss $\kappa=.318$). Our interrater agreement values are considered high comparable to those obtained by \citet{demszky2021measuring} for uptake, and those obtained in widely-used classroom observation protocols such as MQI and the Classroom Assessment Scoring System (CLASS) \citep{pianta2008classroom}. The lower agreement value for \filtered{} indicates that distinguishing funneling vs focusing questions is more subjective than evaluating if a teacher utterance meets the criteria for a follow-up question. This is expected, since the criteria are relatively straightforward and do not require domain expertise.

\begin{table}
\centering
\resizebox{\linewidth}{!}{
\begin{tabular}{@{}ll@{}}
\toprule
\textbf{Example}                                                                                                                                                                                   & Label            \\ \midrule
\begin{tabular}[c]{@{}l@{}}
S:  I disagree with Student A because if you skip count \\ by 100 ten times, that will get you to 1,000. \\
T:  Let’s try it.  You ready?  Let’s \\ start right here with Student F. \\
S:  A hundred.
\end{tabular}                                                                  &  focus \\ \midrule
\begin{tabular}[c]{@{}l@{}}
S:  I first got 32 and then I got 48. \\
T:  And how did you find that? \\
S:  Because I did 16 times two is 32. \\
\end{tabular}                                                                  &  focus \\ \midrule
\begin{tabular}[c]{@{}l@{}}
S:  We did 5 times 2 equals 10. \\
T:  No.  We did 5 times 1 equals 5, darling. \\
S:  Oh, that's to solve the whole –
\end{tabular} & funnel  \\  \midrule
\begin{tabular}[c]{@{}l@{}}
S:  4 minus 2 equals 2. \\
T:  Two and eight tenths.  Does everybody understand? \\
S:  Yes. \\
\end{tabular} & funnel  \\ \midrule
\begin{tabular}[c]{@{}l@{}}
S:  Are we gonna out in the hallway?\\
T:  Yeah.\\
S:  Please.
\end{tabular}                                             & N/A \\ \midrule
\begin{tabular}[c]{@{}l@{}}
S:  I’m not going to [keep it]. \\
T:  Why?  When you’re ready to let me help you, \\ please let me know. \\
S:  [Multiple conversations].\end{tabular}                                             &  N/A  \\ \bottomrule
\end{tabular}
}

\caption{Examples from our annotated data, showing the majority label for each example. \label{tab:annotated_examples}}
\end{table}

\subsection{Educational Outcomes}
\label{ssec:outcomes}

In order to understand the relationship between funneling and focusing and instruction quality and learning outcomes, we leverage variables associated with the original transcript dataset described above, from which we sampled our data. We use classroom observation scores from the MQI coding instrument \citep{hill2008mathematical} for the following items: (1) students provide explanations (scale: not present, low, mid, high), (2) overall student-participation and meaning making and reasoning (scale: not present, low, mid, high), (3) mathematical quality of instruction (5 point scale: low, low/mid, mid, mid/high, high). We chose these items as they relate most closely to questioning patterns and their effect on students discourse. We also use value-added scores, statistical estimates of a teacher’s contribution to student test score growth. Value-added models make statistical adjustments to account for differences in student learning that might result from student background or school-wide factors outside the teacher’s control. Numerous studies in education and economics have shown that value-added scores are an unbiased estimate of teacher impact on student achievement \citep[e.g.][]{chetty2014measuring}. It has also been widely used in teacher evaluation systems around the country.
\section{Proposed Methods}
We use a variety of supervised and unsupervised methods to identify funneling and focusing questions.

\begin{table*}[t!]
\resizebox{\linewidth}{!}{
    \label{crouch}
    \begin{tabular}{  l  p{10.8cm} }
        \toprule
\textbf{Features} 
& \textbf{} \\\midrule
Cognitive verbs 
& understand, think, know, believe, figure out, find out, deduce, remember, imagine, realize, discover \\\hline

Question Words - Unigrams       
& who, what, where, when, why, how, which \\\hline

Question Phrases - Bigrams        
& how many, how do, what is, what else, etc. \\\hline
        \bottomrule
    \end{tabular}
    }
\caption{The list of of our lexical features. We count the appearances of all cognitive verbs and each question word/phrase in an utterance as features to predict funneling and focusing.}
\label{tab:feature_table}
\end{table*}

\paragraph{RoBERTa.} We fine-tune a RoBERTa-based regression model \cite{liu2019roberta} on our annotated data. For our \filtered{} and \unfiltered{} subsets, we trained and evaluated separate models on their respective splits. We performed a small hyperparameter search over the number of epochs, which led to our best model trained over 10 epochs and with the default parameters from the Simple Transformers library \cite{rajapakse2019simpletransformers}.

\paragraph{Forwards-range.} The natural split in the diversity of responses to funneling and focusing utterances led us to adapt \citet{zhang-danescu-niculescu-mizil-2020-balancing}'s forwards-range measure for our task. The forwards-range is an unsupervised measure that quantifies the strength of our expectation of a reply to a given utterance. This measure was used in \citet{zhang-danescu-niculescu-mizil-2020-balancing}'s paper originally to analyze counseling conversations, and here we apply this measure to our dataset.

We use the implementation of the forwards-range from ConvoKit, an open-source toolkit for analyzing conversations \cite{chang-convokit}. ConvoKit transforms each utterance  into a vector representation using TF-IDF re-weighting. Then, to calculate the forwards-range for a given word or phrase, it calculates the weighted average of the vectors for all utterances containing the word/phrase, which \citet{zhang-danescu-niculescu-mizil-2020-balancing} calls the central point. The forwards-range of the word is then calculated as the average cosine distance between the observations containing the word and the central point. 

Before we calculate the forwards-range, we also apply an original pre-processing pipeline to adapt the forwards-range measure to best work in the context of educational data. We apply the following pre-processing pipeline to reduce the vocabulary size and better capture teachers’ rhetorical moves. We first delexicalize all nouns and numbers with “[NOUN]” and “[NUMBER]” tokens. Then, we keep either the last two sentences or the last twenty tokens, whichever one yields the most tokens, following the observation that teachers' questions tend to be at the end of their utterance. We then clean the text by removing punctuation and converting to lowercase. Finally, using the Phrases module of the open-source NLP library Gensim, we find the most common pairs of words in our \unfiltered{} split of the NCTE dataset \cite{rehurek2011gensim}. We use a threshold of $1.0$ to the default Phrases scoring function and a minimum count of 500. The module then joins  the individual words in the bigrams by an underscore character. For example, "okay and how did you do that" becomes “okay and how\_did you do that”. We then apply the ConvoKit framework to our dataset to generate forwards-range scores. 

\paragraph{Length and lexical features.} We also explore the effectiveness of other features in measuring funneling and focusing. We look at (1) length, (2) the count of cognitive verbs, and (3) the count of question words. We calculate length as the number of tokens in a teacher utterance without any pre-processing; this serves as a baseline lexical feature with which to compare performance. In selecting other features, we saw that focusing utterances tended to contain cognitive verbs, which makes sense intuitively since focusing asks students to reflect on their own and/or their classmates' thinking. For the count of cognitive verbs, we source our cognitive verbs from research in cognitive linguistics \cite{cognitive-verbs}. We also include question words after exploratory data analysis, which revealed question words to be predictors of the diversity of responses (e.g., a high range of responses to "why\_did" versus a smaller range of responses for "how\_many"). For question words and phrases, we take the most frequent unigram and bigram question words and phrases present in the NCTE dataset. Table~\ref{tab:feature_table} includes these features.
\section{Experiments and Results}
We evaluate the ability of our models to identify funneling and focusing questions on both the \unfiltered{} and the \filtered{} datasets. We train separate models on each dataset, the idea being that the model trained on the \unfiltered{} set can help identify funneling and focusing questions in "in the wild" -- i.e. in any teacher utterance, while the model trained on the \filtered{} set can help categorize a dataset of questions as funneling or focusing.

The results are shown in Table~\ref{tab:human_correlations}.  We find that the RoBERTa models have a strong positive Spearman correlation with human expert labels both on the \unfiltered{} ($\rho = .761$) and the \filtered{} ($\rho = .443$) sets. Given that the model's score is in a similar range as human agreement\footnote{Human agreement and model scores are not directly comparable. The human agreement values are averaged leave-out estimates across raters (skewed downward). The models’ scores represent correlations with an averaged human score, which smooths over the interrater variance of 3 raters.}, it is unclear if our model has hit a ceiling, or if there is room for improvement above these correlations.

Only few of the unsupervised measures show significant correlations with human judgments. The forwards-range has a significant negative correlation for the \unfiltered{} set ($\rho = -.130$), but it changes to a significant positive correlation for the \filtered{} set ($\rho = .159 $). The positive correlation on the \filtered{} set validates our assumption that focusing questions receive a greater variety of student responses. The negative correlation of the \unfiltered{} suggests that replies to follow-up questions are less varied than other student utterances, which makes intuitive sense, since replies to follow-up questions may reuse words (e.g. ``I think...", ``Yes.") and they tend to stay within the same topic as the teachers' question.

The correlation pattern for length is the opposite as that of the forwards-range, showing a positive correlation with human judgments on the \unfiltered{} set and a negative correlation on the \filtered{} set.  This suggests that overall, teacher utterances containing follow-up questions tend to be longer but that focusing questions tend to be shorter than funneling ones.

As for the other linguistic features, we see a significant positive correlation between the use of ``why'', ``how do'', and ``what else'' on the \filtered{} set, confirming our hypotheses that indicators of open-ended questions are also indicators of focusing. In contrast, the use of ``when'' has a negative correlation with focusing on the \filtered{} set, indicating that that teachers tend to use ``when'' when they expect a normative answer. Interestingly, other question words do not show a significant correlation on the \filtered{} set, indicating that question words in themselves are not strong indicators of funneling and focusing. Question words and cognitive verbs tend to have a positive correlation with humans on the \unfiltered{} set, which is unsurprising, as these features are all indicators of questions. Overall, the trend that we see throughout the unsupervised measures is that there is not enough signal for them to reliably identify funneling and focusing questions.

To measure the practical utility of our models in classroom settings, we also calculated the correlations of our model outputs with educational outcomes (see Section~\ref{ssec:outcomes}). Table~\ref{tab:outcome_results} show the results of this analysis. The observation scores are annotated at the transcript level, so, similar to \cite{demszky2021measuring}, we first mean-aggregate each model’s outputs to yield a model score per transcript. We then use ordinary least squares regression to compute the correlation of the models’ outputs and the outcome scores, controlling for the number of student-teacher exchanges in each transcript. We find that there is a positive linear correlation of the RoBERTa model output scores with all three educational outcome scores for the NCTE dataset. We also find that there is a significant, but weaker correlation between the forwards-range measure and the educational outcomes. 

% Please add the following required packages to your document preamble:
% \usepackage{graphicx}
\begin{table}[t]
\centering
\resizebox{\linewidth}{!}{%
\begin{tabular}{|l|c|c|}
\hline
\multicolumn{1}{|c|}{Models} & \begin{tabular}[c]{@{}c@{}}UNFILTERED \\ (N=2348)\end{tabular}      & \begin{tabular}[c]{@{}c@{}}FILTERED \\ (N=1566)\end{tabular}        \\ \hline
Forwards-range                       & -0.130***         & 0.159***          \\ \hline
Length                               & 0.153***          & -0.149***         \\ \hline
                                     &                   &                   \\ \hline
\multicolumn{1}{|c|}{Question Words} &                   &                   \\ \hline
Who                                  & 0.015             & -0.026            \\ \hline
What                                 & 0.276***          & 0.002             \\ \hline
When                                 & 0.026             & -0.065*           \\ \hline
Where                                & 0.027             & -0.020            \\ \hline
How                                  & 0.189             & -0.036            \\ \hline
Why                                  & 0.188***          & 0.128***          \\ \hline
                                     &                   &                   \\ \hline
How Many                             & 0.065**           & -0.040            \\ \hline
How Do                               & 0.104***          & 0.080**           \\ \hline
What’s                               & 0.051*            & -0.035            \\ \hline
What Else                            & 0.116***          & 0.111***          \\ \hline
                                     &                   &                   \\ \hline
Cognitive Verbs                      & 0.193***          & -0.027            \\ \hline
RoBERTa (unfiltered)                 & \textbf{0.761***} & 0.329***          \\ \hline
RoBERTa (filtered)                   & 0.374***          & \textbf{0.443***} \\ \hline
Interrater correlation       & \begin{tabular}[c]{@{}c@{}}0.619 \\ {[}0.530, 0.694{]}\end{tabular} & \begin{tabular}[c]{@{}c@{}}0.318 \\ {[}0.220, 0.413{]}\end{tabular} \\ \hline
\end{tabular}%
}
\caption{Spearman correlations of model outputs from the supervised RoBERTa model, unsupervised forwards-range model, and word phrase count features with the averages of human labels for question category. Asterisks indicate that the correlation is significant (p-value: *: \textless{}0.05, **\textless{}0.01, ***\textless{}0.001). The brackets for interrater correlation indicate the range of values for 13 raters, where each value represents leave-out correlation for a particular rater.}
\label{tab:human_correlations}
\end{table}

We conduct a similar analysis with value-added scores. Since value-added scores are linked to teachers, we mean-aggregate each models' outputs at the teacher-level. Then, we  compute the linear correlation between each feature and the outcome. The predictions from the RoBERTa model trained on the \filtered{} dataset have a significant correlation with value-added scores, indicating that the measure of funneling and focusing teacher questions captures meaningful variance in students' academic outcomes.

% Please add the following required packages to your document preamble:
% \usepackage{graphicx}
\begin{table*}[t]
\centering
\resizebox{\textwidth}{!}{%
\begin{tabular}{|l|c|c|c|c|}
\hline
 & \begin{tabular}[c]{@{}c@{}}Mathematical Quality \\ of Instruction (MQI5) \\ (N=1657)\end{tabular} & \begin{tabular}[c]{@{}c@{}}Overall Student \\ Participation in \\ Meaning-Making \\ and Reasoning  (N=1657)\end{tabular} & \begin{tabular}[c]{@{}c@{}}Students Provide \\ Explanations \\ (N=1310)\end{tabular} & \begin{tabular}[c]{@{}c@{}}Value-Added \\ Scores (N=304)\end{tabular} \\ \hline
Forwards-range & 0.111*** & 0.209*** & 0.134*** & 0.031 \\ \hline
Length & 0.039 & -0.085* & -0.111*** & -0.096 \\ \hline
 &  &  &  &  \\ \hline
Question Words &  &  &  &  \\ \hline
Who & 0.063*** & 0.008 & 0.009 & -0.361 \\ \hline
How & 0.098*** & 0.001 & 0.003 & -0.101† \\ \hline
What & 0.029 & -0.006 & -0.009 & -0.012 \\ \hline
Where & 0.020 & -0.007 & -0.013 & -0.091 \\ \hline
When & 0.049* & -0.016** & -0.0185† & -0.045 \\ \hline
Why & 0.095*** & 0.050*** & 0.0420*** & -0.03 \\ \hline
 &  &  &  &  \\ \hline
How Many & 0.054** & -0.008 & -0.008 & -0.117* \\ \hline
How Do & 0.076*** & 0.023*** & 0.012 & 0.072 \\ \hline
What’s & 0.001 & -0.018** & -0.024** & 0.027 \\ \hline
What Else & -0.031† & 0.007 & 0.013 & 0.005 \\ \hline
 &  &  &  &  \\ \hline
Cognitive Verbs & 0.105*** & 0.070* & 0.081** & 0.003 \\ \hline
 &  &  &  &  \\ \hline
RoBERTa (unfiltered) & 0.315*** & 0.270*** & 0.350*** & 0.098† \\ \hline
RoBERTa (filtered) & 0.067** & 0.388*** & 0.376*** & \textbf{0.124*} \\ \hline
\end{tabular}%
}
\caption{Standardized coefficients showing the correlation between each measure, including RoBERTa, forwards-range, length and our lexical features, and the outcomes from the NCTE dataset. Each co-efficient comes from its own linear model, with the number of student-teacher exchanges in each transcript as a control variable (p-value: $\dagger$: \textless{}0.1, *: \textless{}0.05, **\textless{}0.01, ***\textless{}0.001).}
\label{tab:outcome_results}
\end{table*}

% Please add the following required packages to your document preamble:
% \usepackage{multirow}
% \usepackage{graphicx}
% \usepackage[table,xcdraw]{xcolor}
% If you use beamer only pass "xcolor=table" option, i.e. \documentclass[xcolor=table]{beamer}
\begin{table*}[t]
\centering
\resizebox{\textwidth}{!}{%
\begin{tabular}{|l|c|cc|}
\hline
 &  & \multicolumn{2}{c|}{Models} \\ \cline{3-4} 
\multirow{-2}{*}{Example Exchange} & \multirow{-2}{*}{Human label} & \multicolumn{1}{c|}{RoBERTa} & \begin{tabular}[c]{@{}c@{}}Forwards-\\ Range\end{tabular} \\ \hline
\begin{tabular}[c]{@{}l@{}}Student:  To see how many twirls.\\ Teacher:  How many – what do you mean?\\ Student:  How many [inaudible] there are.\end{tabular} &  & \multicolumn{1}{c|}{\cellcolor[HTML]{6AA84F}focusing} & \cellcolor[HTML]{EA9999}funneling \\ \cline{1-1} \cline{3-4} 
\begin{tabular}[c]{@{}l@{}}Student:  H-U-N-D-E-R-E-T-H?\\ Teacher:  Sh.  Don’t steal his knowledge.  And Student K?\\ Student:  Thousandths.\end{tabular} & \multirow{-2}{*}{Focusing} & \multicolumn{1}{c|}{\cellcolor[HTML]{EA9999}funneling} & \cellcolor[HTML]{6AA84F}focusing \\ \hline
\begin{tabular}[c]{@{}l@{}}Student:  I put about, about is that about?\\ Teacher:  It is close.\\ Student:  What’s the about sign?\end{tabular} &  & \multicolumn{1}{c|}{\cellcolor[HTML]{6AA84F}{\color[HTML]{000000} funneling}} & \cellcolor[HTML]{6AA84F}funneling \\ \cline{1-1} \cline{3-4} 
\begin{tabular}[c]{@{}l@{}}Student:  Three twelfths also equals one quarter.\\ Teacher:  Yes, it does, and we’ll talk about that in another lesson, okay? \\ Number 10.  Cover Shape B with –\\ Student:  Hexagons.\end{tabular} & \multirow{-2}{*}{Funneling} & \multicolumn{1}{c|}{\cellcolor[HTML]{6AA84F}funneling} & \cellcolor[HTML]{6AA84F}funneling \\ \hline
\end{tabular}%
}
\caption{Example model predictions from the forwards-range and our RoBERTa model fine-tuned on the \filtered{} set. Correct predictions are in green, and incorrect predictions are in red.}
\label{tab:example_predictions}
\end{table*}

\section{Qualitative Analysis of Model Outputs}

To better understand the performance of our models, we analyzed the predictions of our RoBERTa model fine-tuned on the \filtered{} set and the forwards-range. Here, we choose to analyze performance on the \filtered{} set to better understand the performance of our models in specifically distinguishing between funneling and focusing, rather than including the \unfiltered{} set for the related but easier task of identifying if the teacher prompted the student. Some selected examples that we examined are shown in Table~\ref{tab:example_predictions}. For utterances without question phrases, the RoBERTa model and forwards-range model perform better than stand-alone question phrase features, as shown in the last example in Table~\ref{tab:example_predictions}. Many funneling teacher utterances do not actually include question words, but rather prompt the student to finish the teacher's sentence. The complexity of these classes of sentences, covering a wide range of topics with unique vocabulary tokens, motivates the use of our forwards-range and RoBERTa models, which are able to correctly classify these examples.

The RoBERTa model was also able to classify more complex examples that include several different question phrases. For instance, the first example in the table, includes the question phrase “how many”, which correlates with funneling. But then the teacher also asks the student about their thinking, asking “what do you mean” by that, which makes the utterance an example of focusing. This suggests that the RoBERTa model is able to account for contextual factors and weigh the importance of different question phrases. On the other hand, the forwards-range predicted this example as “funneling”, which shows one of its weaknesses as a bag-of-words model that lacks context.

One area of improvement across all the models we found through manual inspection is a class of focusing examples where the teacher calls on students to reflect on other students’ contributions. For example, if a teacher asks a student Student B, “Is Student A correct?”, this is a closed-ended question that could be interpreted as funneling, but it is focusing since the student is reflecting on the thinking of another student. The second entry in Table~\ref{tab:example_predictions} also illustrates this, as the teacher asks a follow up to one student after receiving an answer from a different student. The RoBERTa model predicts this as funneling, likely because the utterance ends with a short, closed-ended question. However, this is an example of focusing, as the teacher calls on Student K to reflect on the previous student’s thinking. The forwards-range predicts this example as focusing, but we do not believe that, as a bag of words model, the forwards-range actually captures the nuance of this example. It instead might be unsure of the expected reply and default predicts focus since the majority of its scores are clustered around relatively high forwards-range scores.

\section{Conclusion and Future Work}
We propose several approaches for computationally measuring funneling and focusing, an important aspect of classroom discourse, and evaluate their strengths and weaknesses. Our supervised approach using the fine-tuned RoBERTa model has the strongest linear correlation of the models we tested with human expert ratings for funneling and focusing; it similarly had the strongest correlations of the models with educational outcomes. This shows the potential of the RoBERTa model to be used in future feedback and professional development tools for teachers. 

Still, our unsupervised measures show significant correlations with the expert labels for funneling and focusing, as well as with educational outcomes. This provides a foundation for future work combining different unsupervised approaches to build a robust measure of funneling and focusing. Other paths for future NLP work include using probing and attention weights to better understand the predictions of the RoBERTa model, improving the supervised approaches via an extensive hyperparameter search and by exploring models beyond RoBERTa, and importantly, improving and testing the generalizability of this measure to other classrooms and domains. 

In education, there is potential for future work in exploring how this measure can best support instruction and learning outcomes for students across different educational settings. One possible avenue for this is examining discipline-specific ways of identifying focusing or funneling to provide more fine-grained feedback to teachers. Another is investigating the extent to which it is helpful that teachers know quantitatively in feedback they receive how much they are focusing versus funneling, or if there's a qualitative element about focusing and funneling that could similarly be helpful to teachers if provided in feedback.
\section*{Acknowledgments}
We thank the anonymous reviewers for their helpful feedback. We are also grateful for the generous support of the Stanford CURIS program (to S. Alic) and the Melvin and Joan Lane Stanford Graduate Fellowship (to D. Demszky). The research reported here was supported in part by the Institute of Education Sciences, U.S. Department of Education (Grant R305C090023) to the President and Fellows of Harvard College to support the National Center for Teacher Effectiveness. The opinions expressed are those of the authors and do not represent views of the Institute or the U.S. Department of Education.

% Entries for the entire Anthology, followed by custom entries
\bibliography{custom}
\bibliographystyle{acl_natbib}

% This is an appendix.

\end{document}